\documentstyle[12pt]{article}
\def\be {\begin{equation}}
\def\ee {\end{equation}}
\def\ba {\begin{eqnarray}}
\def\ea {\end{eqnarray}}

\def\bi {\begin{itemize}}
\def\ei {\end{itemize}}
\begin{document}
\def\bea{\begin{eqnarray}}
\def\eea{\end{eqnarray}}
\title{\bf {Area Spectrum of Near Extremal Black Branes
from Quasi-normal Modes }}
 \author{M.R. Setare  \footnote{E-mail: rezakord@ipm.ir}
  \\{Physics Dept. Inst. for Studies in Theo. Physics and
Mathematics(IPM)}\\
{P. O. Box 19395-5531, Tehran, IRAN }}
\date{\small{}}

\maketitle
\begin{abstract}
Motivated by the recent interest in quantization of black hole
area spectrum, we consider the area spectrum of near extremal
black $3-$branes. Based on the proposal by Bekenstein and others
that the black hole area spectrum is discrete and equally spaced,
we implement Kunstatter's method to derive the area spectrum for
the near extremal black $3-$branes. The result for the  area of
event horizon although discrete, is not equally spaced.
 \end{abstract}
\newpage

 \section{Introduction}
Dynamical properties of a thermal gauge theory are encoded in its
Green's functions. In the context of AdS/CFT \cite{mal},
Minkowski space Green's functions can be computed from gravity
using the recipe given in \cite{stari1}. Unfortunately, for a
non-extremal background, only approximate expressions for the
correlators are usually obtained. Even thought the retarded
Green's function in $4d$ cannot be found explicitly, the location
of its singularities can be determined precisely.  As shown in
\cite{stari1}, this amounts to finding the quasinormal
frequencies of dilaton's fluctuation in the dual near-extremal
black brane background as functions of the spatial momentum.
 The possibility of
a connection between the quasinormal frequencies of black holes
and the quantum properties of the entropy spectrum was first
observed by Bekenstein \cite{bek3}, and further developed by Hod
\cite{hod}. Bekenstein noted that Bohr's correspondence principle
implies that frequencies characterizing transitions between
energy levels of a quantum black hole at large quantum numbers
correspond to the black hole's classical oscillation
(quasinormal) frequencies (see also \cite{{kok1},{nol}}). In
particular, Hod proposed that the real part of the quasinormal
frequencies, in the infinite damping limit (i.e. the
$n\rightarrow\infty$ limit), might be related via Bohr's
correspondence principle to the fundamental quanta of mass and
angular momentum (see also
\cite{bek4}--\hspace{-0.1ex}\cite{pad3}).
\\In asymptotically flat spacetimes the idea of QNMs started with the
work of Regge and Wheeler \cite{reggeW} where the stability of a
Schwarzschild black hole was tested, and were first numerically
computed by Chandrasekhar and Detweiler several years later
\cite{Chandra1}. Recently, there has been considerable interest
in studing the quasinormal modes in different contexts: in
AdS/CFT duality conjecture
\cite{Horowitz-Hubeny}--\hspace{-0.1ex}\cite{carli}, when
considering thermodynamic properties of black holes in loop
quantum gravity \cite{dry}--\hspace{-0.1ex}\cite{hon}, in the
context of possible connection with critical collapse
\cite{Horowitz-Hubeny,kim, BHCC}, also when considering the area
spectrum of black holes {\cite{set1}-\cite{das}}. Recently it has
been observed that the quasinormal modes can play a fundamental
role in Loop Quantum Gravity \cite{dry}. Dreyer showed that in
order to have consistence between the Bekenstein-Hawking entropy
calculation and QNM frequencies, one had to assume that the
minimum of $j$ of the spin network piercing the horizon and
contributing significantly to the entropy had to be $j=1$. Whit
this choice, the resulting Immrizi parameter would be given by
$\gamma=\frac{\ln3}{2\pi\pi\sqrt{2}}$. He suggested that if the
gauge group of the theory were changed from $SU(2)$ to $SO(3)$
then this requirement would be immediately satisfied.
\\ For the Schwarzschild black hole in four dimensions, the
asymptotic real part of the gravitational quasinormal frequencies
is of the form $\omega=T_{H}ln 3$ where $T_H$ is the Hawking
temperature \cite{mot}. The suggestion of Hod was to identify
$\hbar \omega$ with the fundamental quantum of mass $\Delta M$.
This identification immediately leads to an area spacing of the
form $\Delta A=4\hbar ln3$. An elegant approach, for the
schwarzschild black hole in $d-$dimensions, based on analytic
continuation and computation of the monodromy of the perturbation
was proposed in \cite{motnit}, (to see more recent works refer to
\cite{recent}).
\\
In the present paper  we extend directly the Kunstatter's
approach \cite{kun} to determine mass and area spectrum of the
near extremal black $3-$branes . According to this approach, an
adiabatic invariant $I=\int {dE\over \omega(E)}$, where $E$ is
the energy of system and $\omega(E)$ is the vibrational
frequency, has an equally spaced spectrum, i.e. $I\approx
n\hbar$, applying the Bohr-Sommerfeld quantization at the large
$n$ limit.

\section{Coincident D-3 Branes}
We consider now the background (in the string frame) of a black
hole describing a number of coinciding D-3 branes \cite{kel, kir}
\be ds^{2}=H^{-1/2}(r)[-f(r) dt^{2}+\sum_{i=1}^{3}(dx^{i})
^{2}]+H^{1/2}[ f^{-1}(r)dr^{2}+r^{2}d\Omega_{5}^{2}], \label{met}
\ee where \be H(r)=1+\frac{l^{4}}{r^{4}}, \hspace{1cm}
f(r)=1-\frac{r_{0}^{4}}{r^{4}}, \label{hfeq} \ee and
$d\Omega_{5}^{2}$ is the metric of a unit $5-$dimensional sphere.
The horizon is located at $r=r_{0}$ and the extremality is
achieved in the limit $r_{0}\rightarrow 0$. For $l$ much larger
than the string scale $\sqrt{\alpha'}$, the entire $3-$brane
geometry has small curvatures every where and is appropriately
described by the supergravity approximation to type $IIB$ string
theory \cite{kel}. The requirement $l\gg\sqrt{\alpha'}$
translates into the language of $U(N)$ SYM theory on $N$
coincident $D-3$branes. To this end it is convenient to equate
the ADM  tension of the extremal $3-$brane classical solution to
$N$ times the tension of a single $D3-$ brane. Then one can find
\cite{gab} \be
\frac{2l^{4}\Omega_{5}}{k^{2}}=N\frac{\sqrt{\pi}}{k},
\label{const}\ee where $\Omega_{5}=\pi^{3}$ is the volume of a
unit $5-$sphere, and $k=\sqrt{8\pi G}$ is the $10-$dimensional
gravitational constant. Therefore \be
l^{4}=\frac{kN}{2\pi^{5/2}}. \label{leq} \ee In the other hand we
have \be k=8\pi^{7/2}g_{s}\alpha'^{2}, \label{string} \ee where
$g_{s}$ is the string coupling, then we obtain \be  l^{4}=4\pi N
g_{s}\alpha'^{2}. \label{coupling} \ee The parameters $l$ and
$r_{0}$ are related to the ADM mass in the following way \be
M=\frac{\Omega_{5}V_{3}}{2k_{10}^{2}}(5r_{0}^{4}+4l^{4}).\label{mass1}
\ee Now we would like to consider the near-extremal $3-$brane
geometry. In the near-horizon region, $r\ll l$, we may replace
$H(r)$ by $\frac{l^{4}}{r^{4}}$. The resulting metric is as
following \be
ds^{2}=\frac{r^{2}}{l^{2}}[-(1-\frac{r_{0}^{4}}{r^{4}})dt^{2}+d\overrightarrow{x}^{2}]
+\frac{l^{2}}{r^{2}}(1-\frac{r_{0}^{4}}{r^{4}})
^{-1}dr^{2}+l^{2}d\Omega_{5}^{2}.\label{near} \ee The above
metric is a product of $S^{5}$ with a certain limit of the
Schwarzschild black hole in $AdS_{5}$. The $8-$dimensional area
of the horizon can be read off from metric (\ref{near}). If the
spatial volume of the $D3-$brane is taken to be $V_{3}$, then we
find \be A_{h}=( \frac{r_{0}}{l})
^{3}V_{3}l^{5}\Omega_{5}=\pi^{6}l^{8}T^{3}V_{3}, \label{hori} \ee
where $T$ is a temperature \be T=\frac{r_{0}}{\pi l^{2}}.
\label{temp} \ee Using (\ref{leq}) we arrive at the
Bekenstein-Hawking entropy \cite{gab} \be S_{BH}=\frac{2\pi
A_{h}}{k^{2}}=\frac{\pi^{2}N^{2}V_{3}T^{3}}{2}.\label{entro} \ee
\section{Quasinormal Modes and Area Spectrum}
Given a system with energy $E$ and vibrational frequency
$\omega(E)$, one can show that the quantity \be I=\int
\frac{dE}{\omega(E)} \label{adia1} \ee where $dE=dM$, is an
adiabatic invariant \cite{kun} and as already mentioned in the
Introduction, via Bohr-Sommerfeld quantization has an equally
spaced spectrum in the large $n$ limit \be I \approx n\hbar
\hspace{1ex}. \label{bohr} \ee The large $n$ asymptotic behavior
of quasinormal frequencies given by following
expression\cite{star} \be \omega_{n}^{\pm}=\omega_{0}^{\pm}\pm
2\pi T n(1\mp i),\label{quasi} \ee where \be \omega_{0}^{\pm}=\pi
T \lambda_{0}^{\pm}, \label{omga}  \ee where
$\lambda_{0}^{\pm}\approx \pm 1.2139-0.7775i$. Now by taking
$\omega_{R}$ as \be \omega_{R}=\pm \pi T (2n+1.2139),
\label{real1} \ee then by substituting Eq.(\ref{temp}) we get \be
\omega_{R}=\frac{\pm r_{0}}{l^{2}}(2n+1.2139).\label{real2} \ee
By taking $M$ as given by Eq.(\ref{mass1}) and substituting
Eq.(\ref{coupling}), we obtain \be
M=\frac{V_{3}}{128\pi^{4}\alpha'^{4}g_{s}^{2}}(5r_{0}^{4}+16\pi N
g_{s}\alpha'^{2}).\label{mass2} \ee Then, the parameter $r_{0}$ is
given by \be r_{0}=(aM-b) ^{1/4}, \label{param} \ee where \be
a=\frac{128 \pi^{4}\alpha'^{4}g_{s}^{2}}{5V_{3}}, \hspace{1cm}
b=\frac{16\pi N \alpha'^{2}g_{s}}{5}. \label{abeq} \ee Now by
taking $\omega_{R}$ as given by expression (\ref{real2}) and
substituting Eqs.(\ref{param},\ref{coupling}), we get \be
\omega_{R}=\frac{\pm (aM-b) ^{1/4}}{2\alpha' \sqrt{\pi N
g_{s}}}(2n+1.2139).\label{real3} \ee Thus, the adiabatically
invariant integral (\ref{adia1}) is written as \be I=\frac{\pm 2
\alpha'\sqrt{\pi N g_{s}}  }{(2n+1.2139)}\int\frac{dM}{(aM-b)
^{1/4}},\label{adia2} \ee and after integration, we obtain \be
I=\frac{5V_{3}\sqrt{N}}{48\pi^{7/2}\alpha'^{3}g_{s}^{3/2}(2n+1.2139)}(aM-b)^{3/4}.
\label{adia3} \ee Now using Eqs.(\ref{temp},\ref{param}) we can
rewrite the area of the horizon Eq.(\ref{hori}) as \be
A_{h}=\pi^{3}l^{2}V_{3}r_{0}^{3}=\pi^{3}l^{2}V_{3}(aM-b)^{3/4}.\label{hori2}
\ee Using Eq.(\ref{adia3}) the area is given by following
expression \be
A_{h}=\frac{3(2n+1.2139)}{160}\pi^{7}\alpha'^{4}g_{s}^{2}I=\frac{3}{160}\pi^{7}
\alpha'^{4}g_{s}^{2}(2n+1.2139)n\hbar. \label{hori3} \ee It is
obvious that the area spectrum, although discrete, is not
equivalently spaced. The  quasinormal frequencies  which given by
Eq.(\ref{quasi}) was later generalized in the  paper by Nunez and
Starinets \cite{stari2}, Eqs. (3.22-3.23), but all these formulas
are asymptotics rather than exact results. However, for vector
perturbations (and spatial momentum on the brane equal to zero)
the spectrum turns out to be exact and given by following
relation \be \omega_n=n(1-i), \hspace{1cm} n=0,1,...
\label{exact}\ee therefore, the integral Eq.(\ref{adia1}) yields
\be I=\frac{M}{n}, \label{massad}\ee and by equating expressions
(\ref{massad}) and (\ref{bohr}), we get \be M=n^{2}\hbar.
\label{massspec} \ee It is obvious that the mass spectrum of
$3-$black brane is quantize.

\section{Conclusion}
The possibility of a connection between quasinormal modes and the
area spectrum of black holes has been actively pursued over the
past year. Many examples have been studied, and progress has been
made towards a general understanding of this connection.
Bekenestein's  idea for quantizing a black hole is based on the
fact that its horizon area, in the nonextremal case, behaves as a
classical adiabatic invariant. It is interesting to investigate
how near extremal black $3-$branes would be quantized. Discrete
spectra arise in quantum mechanics in the presence of a
periodicity  in the classical system, which in turn leads to the
existence of an adiabatic invariant  or action variable.
Boher-Somerfeld quantization implies that this adiabatic
invariant has an equally spaced spectrum in the semi-classical
limit. Kunstatter showed that this approach works for the
 Schwarzschild black holes in any dimension, giving asymptotically equally
 spaced areas, previously we have showed generalization to  non-rotating
 BTZ, extremal Reissner-Nordstr\"om, near extremal Schwarzschild-de
 Sitter Kerr and extremal Kerr black holes
 \cite{{set1},{set2},{set3},{set4}}
  is also successful. In this article we have
considered the near extremal black $3-$branes. Using the results
for highly damped quasi-normal modes Eq.(14), we obtained the area
 spectrum of event horizon Eq.(25), which is
obvious that the area spectrum, although discrete, is not
equivalently spaced. Using the generalized form of quasinormal
modes Eq.(26), we obtained mass spectrum of near extremal
$3-$black brane as Eq.(28). Similar situation occur for BTZ black
hole\cite{set1} which the  area of event horizon is not equally
spaced, in contrast with area spectrum of black hole in higher
dimension, although the mass spectrum is equally spaced.

  \vspace{3mm}

\section*{Acknowledgement }
I would like to thank Prof. Andrei O. Starinets which introduced
me the references [43,44,46].

\end{document}